\documentclass{aa}  
\usepackage[]{natbib}
\bibpunct{(}{)}{;}{a}{}{,} % to follow the A&A style
\usepackage{caption}
\usepackage{subcaption}
\usepackage{amsmath}
\usepackage{graphicx}
\usepackage{txfonts}
\usepackage[breaklinks=true,colorlinks=true,urlcolor=blue,citecolor=blue,linkcolor=blue]{hyperref}

\begin{document}
 \title{Solar off-limb emission of the \ion{O}{I}  7772~\AA\ line}

  % \subtitle{let's do this}

   \author{H. Pazira
          \inst{1}
          \and
          D. Kiselman\inst{1}
          \and
          J. Leenaarts\inst{1}
          }

   \institute{Institute for Solar physics, Department of Astronomy, Stockholm University, AlbaNova University Centre, SE-106 91 Stockholm, Sweden
              \\
              \email{hiva.pazira@astro.su.se}
           %  \thanks{}
             }

   \date{Received September --, --; accepted --, --}

% \abstract{}{}{}{}{} 
% 5 {} token are mandatory
 
  \abstract
   {}
   {The aim of this paper is to understand the formation of the \ion{O}{I}  line at 7772~\AA\ in the solar chromosphere.}
   {We used SST/CRISP observations to observe \ion{O}{I}  7772~\AA\ in several
     places around the solar limb. We compared the observations with
     synthetic spectra calculated with the RH code in the
     one-dimension spherical geometry mode. New accurate hydrogen collisional rates were included for the RH calculations.}
   {The observations reveal a dark gap in the lower chromosphere,
     which is caused by variations in the line opacity as shown by our
     models. The lower level of the 7772~\AA\ transition is populated by a downward cascade from the continuum. We study the effect of Lyman-$\beta$ pumping and hydrogen collisions between the triplet and quintet system in \ion{O}{I} . Both have a small but non-negligible influence on the line intensity.}
   {}


   \maketitle

%________________________________________________________________
\section{Introduction}
The infrared (IR) triplet lines of \ion{O}{I}  (7771.9, 7774.2 and 7775.4~\AA) are
the most studied spectral lines of neutral oxygen in solar-type stars.
It has been clear since the work of \citet{1968SoPh....5..260A} 
that these lines are formed under non-local thermodynamic equilibrium (non-LTE) conditions in the solar photosphere.
The photospheric line formation has been investigated in many studies
in a stellar context with the aim of gathering oxygen abundance measurements, for example \citet{2016MNRAS.455.3735A} and references in there.

An early off-limb observation of the IR triplet was made 
in 1928 by W.H. Babcock \citep{1957sun..book.....A}, thus showing that there is some chromospheric emission from
it; something noted also by \citep{1968ApJS...17....1P}. 

Since there are a limited number of chromospheric lines which can be observed with
ground-based instruments, it is useful to try to better understand
the formation of the \ion{O}{I}  7772~\AA\ emission as a chromospheric
diagnostic. This requires detailed modelling of the line-formation processes.
In the solar chromosphere, the density drops and the collisional
rates decrease as compared to the photosphere. Thus strong non-LTE effects are expected for the off-limb emission. 

\citet{1962ApJ...135..500A} and \citet{1995ApJ...438..491A} used the
eclipse flash spectrum to look for non-LTE effects. 
The coincidence of the Ly$\beta$ line and the \ion{O}{I}  1027~\AA\ resonance
lines \citep{1982ApJ...259..869S} has stimulated interest because the
oxygen line is expected to be pumped by Ly$\beta$.  The modelling efforts of
\citet{1993ApJ...402..344C} were primarily directed towards the UV resonance lines and
\citet{2015ApJ...813...34L} concentrated on the 1356~\AA\ intercombination line.

\citet{1999ApJ...518L.131P} made spectroscopic off-limb observations
(from 1500 km to 5000 km above the limb) of the IR
triplet and the IR doublet at 8445~\AA. From the intensity ratios he
concluded that the lines are formed via electron collisional excitation, provided the emission originates from optically thin gas.

\citet{1987ApJ...313..463B} use the 7772 ~\AA\ line to study a quiescent
prominence with a seven-level model atom and find that the line is
temperature sensitive for temperatures around 7000~K because of
collisional excitation from the ground state. 

The goal of this paper is to improve the understanding of the line-formation processes of \ion{O}{I} 
7772 ~\AA\ in the solar chromosphere.
Our strategy is to use an imaging spectrometer to acquire very high-spatial-resolution observations of the off-limb
emission and compare those to the results of non-LTE spectral modelling in spherical geometry. 

In Sect.~\ref{Sec:obs} we describe the observations, in
Sect.~\ref{Sect:obs_res} we present our observational results, and in
Sect.~\ref{Sect:mod} we outline the simulation setup. Sect.~\ref{sect:comp}
discusses the comparison between simulations and observations. In
Sect.~\ref{Sect:understand} we explain the line formation processes
and their driving mechanisms.

\section{Observations}
\label{Sec:obs}
The data used in this work were recorded during the period of October 19-21, 2013 using the CRisp Imaging SpectroPolarimeter 
\citep[CRISP,][]{2008ApJ...689L..69S} at the Swedish 1-m Solar Telescope \citep[SST,][]{2003SPIE.4853..341S}.
CRISP takes quasi-simultaneous images of the Sun at different wavelengths. 

The image scale at 7772 \AA\  is 0\farcs 061 per pixel which results in a field of view of 62\arcsec. 
The frame rate was 36.5\,Hz, which gives an exposure time per frame of 17\,ms.

\begin{figure}
\includegraphics[width=8.8cm]{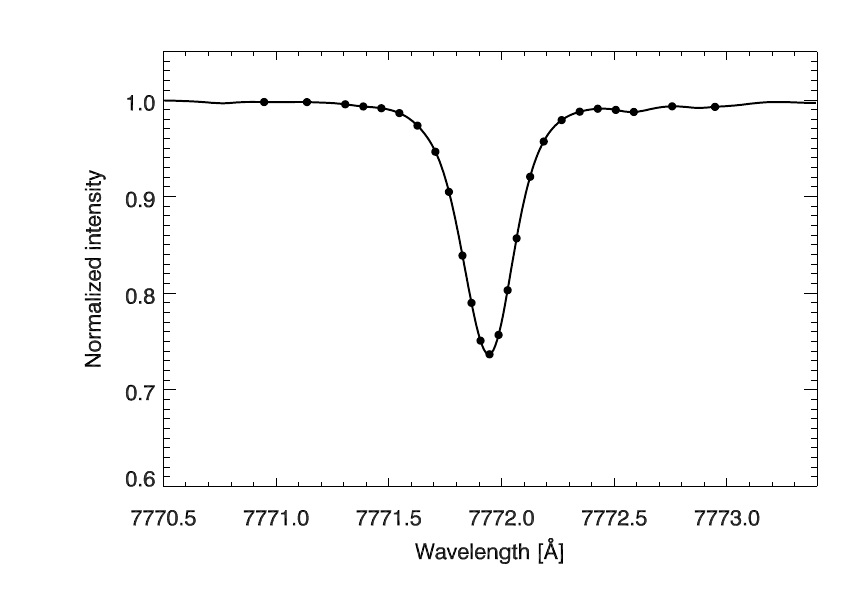}
\caption{Average line profile of \ion{O}{I}  7772 \AA\ from the FTS atlas \citep{1984SoPh...90..205N}. The symbols represent the positions that we sampled with SST/CRISP. }
\label{fig:lnprf}
\end{figure}

These data are non-polarimetric, that is, the liquid-crystal retarders were not cycled, and the scanning sequence consisted of 25 line positions (Fig.~\ref{fig:lnprf}) giving a cadence of about four seconds. 
This is short enough to observe high-speed changes in the solar chromosphere.
%DH Not so interesting The FWHM of  pre--filter is about $77$ m \AA\ (anywhere else except pipeline paper?s) at 7772\,\AA\. 

The CRISPRED pipeline \citep{2015A&A...573A..40D} was used for the data reduction. 
Image reconstruction was made with MOMFBD \citep{2005SoPh..228..191V}.
In the near infrared, backscattered light in the CCD produces an
imprint of an electronic circuit pattern in the image. The procedure
for correcting for this effect \citep{2013A&A...556A.115D} was
included in the reduction pipeline. Still, some traces of this pattern are visible on parts of the reduced frames where the solar signal is weak.

\begin{figure}
\includegraphics[width=8.8cm]{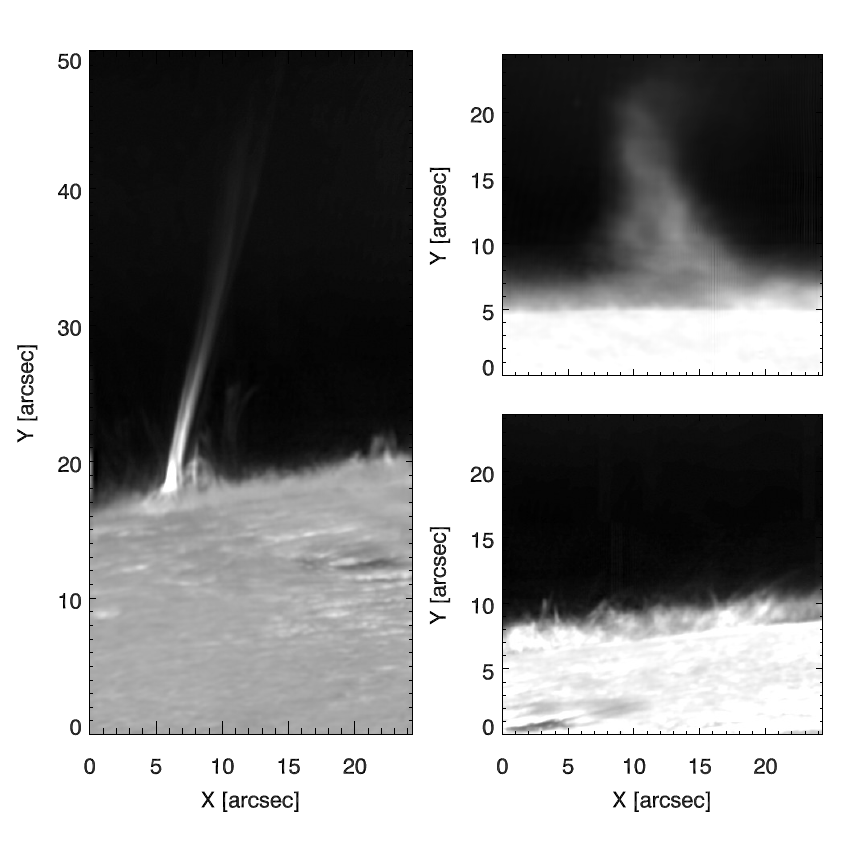}
\caption{ Examples of chromospheric features that can be observed off-limb  in \ion{O}{I}  7772 \AA ~line core , \textit{(left)} surge, \textit{(top right)} a weak prominence
  \textit{(bottom right)} active region spicules. Radial enhancement
  of the intensity has been applied to the images to increase the visibility of off-limb structures. The two right-hand images have also been brightness clipped to further enhance the visibility.}
\label{fig:possib}
\end{figure}

\begin{figure*}
\centering
\includegraphics[height=6cm]{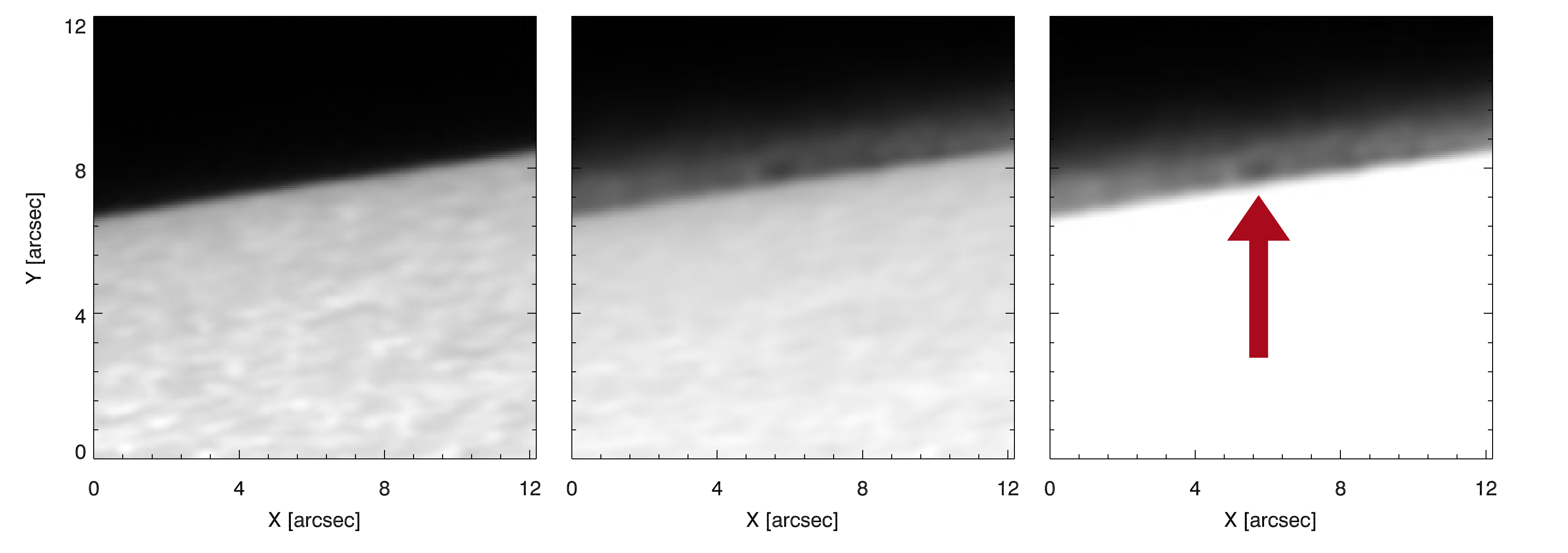}
\caption{Quiet solar limb imaged  in the continuum close to \ion{O}{I}  7772~\AA\ line and in its line core. Left
  panel: continuum image (810 m\AA\ from the line core). Centre panel:
  line core image. Right panel: line core image with intensity scaling
  to better show the off-limb emission. These images have {\em not} undergone any radial enhancement as in Fig.~\ref{fig:possib}.
  The red
  arrow indicates a point on the limb where the dark gap is
  strong and where the radial intensity profile used to compare with
  modelling is extracted.}
\label{fig:qs}
\end{figure*}
%______________________________________________________________
\section{Observational results}
\label{Sect:obs_res}
Off-limb emission in the line core of \ion{O}{I} 7772~\AA\ is discernible around the whole limb. It is stronger in and close to active regions. 
Figure~\ref{fig:possib} illustrates different structures that can be
seen at this wavelength: A surge, spicules close to an active region, and a quiet
prominence. All these off-limb images have been enhanced by a radial
filter similar to that applied by \citet{2014ApJ...792L..15P}.

In order to investigate the origin of the emission and its diagnostic
potential, in the following we concentrate on an image of the
quiet solar limb shown in Fig.~\ref{fig:qs}. The left panel
shows the continuum image at 810 m\AA\ from the line centre. The central panel is the line
core image and the right panel is the same image scaled to highlight
the off-limb emission (while saturating the disc).

The line-core images of Fig.~\ref{fig:qs} show a dark gap between the limb and a
point of maximum intensity of the off-limb emission. This dark gap
can be seen in line-core images almost everywhere along the limb.

We feel confident that the dark gaps are not caused by instrumental
effects nor are they artifacts from the image reconstruction. They can
be seen in non-reconstructed but flat-fielded images. Dark gaps appear at different locations along the
solar limb whenever the seeing is sufficiently good. 

For the quantitative analysis in the following, we have
chosen a part of the limb where the dark gap is most obvious, where
there are no superpositions from spicules, and there is no
visible circuit pattern caused by detector back-scattering. The red
arrow in Fig.~\ref{fig:qs} indicates its position.

%______________________________________________________________
 \section{Modelling}
\label{Sect:mod}
\subsection{Radiative transfer code}
For modelling the off-limb emission, we used the RH code \citep{2001ApJ...557..389U} in its spherical geometry version. This code can solve non-LTE radiative transfer problems and can treat spectral lines using partial redistribution. It has also the possibility to solve the radiative transfer calculations for several elements simultaneously. This is relevant in the case of oxygen which has been suspected for a long time to be affected by hydrogen due to the coincidence of wavelength between the \ion{O}{I} resonance line at 1025~\AA\ and Ly$\beta$,
 \citep[see][and references in that paper]{2015ApJ...813...34L}.
 
Therefore hydrogen was treated in non-LTE simultaneously with
oxygen. We discuss the effects of the coupling to hydrogen on the formation of
\ion{O}{I} 7772 \AA\ in Section~\ref{sec:heff}. After some experimenting,
we decided to also treat silicon in non-LTE because the Si\,{\sc i}
continuum is an important background opacity source at relevant
ultraviolet wavelengths.

The spherical geometry version of RH solves the radiative transfer
equations along various rays and synthesises the spectrum along each
ray. The rays can pass outside the nominal solar limb. This possibility is used here to study off-limb emission.
\subsection{Atomic models}
\begin{figure*} [t]
\includegraphics{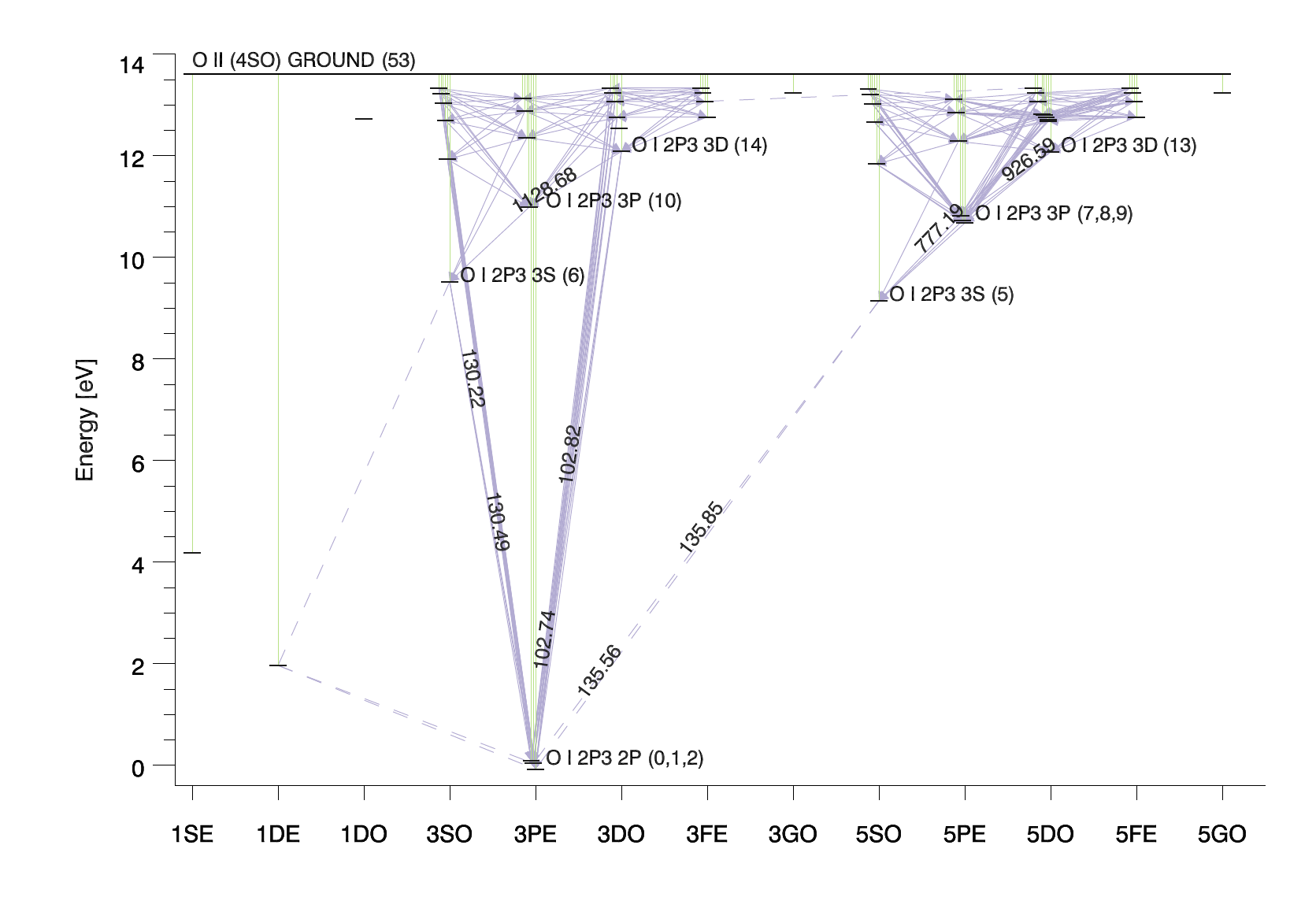}
\caption{Grotrian diagram of the 54-level oxygen model atom used in this work. The key levels important in the discussion of the 7772~\AA\ off-limb emission are labelled with arabic numerals between brackets after the term label.}
\label{fig:term}
\end{figure*} 

The oxygen atomic model used here is the 54-level atom with 258
transitions of \citet{2009A&A...500.1221F}. In this atomic model,
electron collisions for several transitions are from the calculations
by \citet{2007A&A...462..781B}. Of concern are inelastic collisions
with neutral hydrogen which in the model of
\citet{2009A&A...500.1221F} are coming from calculations by
\citet{1984A&A...130..319S} which in turn were based on a generalisation of the
approximation of \citet{1969ZPhy..228...99D} for allowed
transitions. In our model we replaced these rates with
newly calculated ones kindly provided by Dr. P. Barklem (private
communication). These calculations were made using the methods of
Linear Combination of Atomic Orbitals
\citep[LCAO]{2016PhRvA..93d2705B}. They are used for all transitions
between levels in the range 1-20.The atomic levels and the radiative transitions are illustrated in Fig.~\ref{fig:term}.

For the other active species we use the model atoms that come with the
RH distribution. The \ion{H}{I} atom has five levels plus continuum and \ion{Si}{I}\
is represented by 23 levels plus continuum.

\subsection{Atmospheric model}
\begin{figure}[ t]
\includegraphics{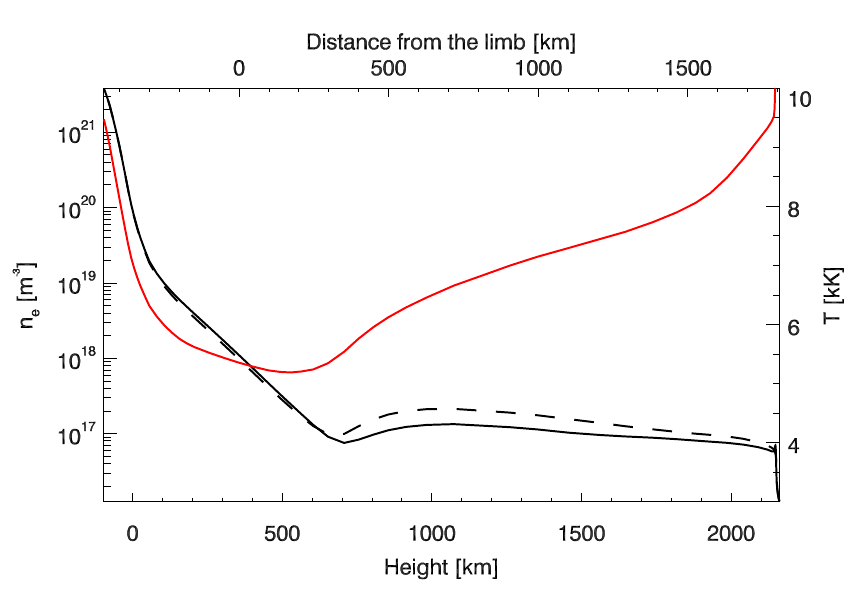}
\caption{ The temperature (red) and the electron number density for the FAL-C
 atmospheric model. The dashed line represents the resulting electron number
 density of the final solution for the Full model as described in the text.}
\label{fig:FAL_C}
\end{figure}

As the baseline atmospheric model we used FAL-C
\citep{1993ApJ...406..319F}. FAL-C is a one-dimensional (1D) semi-empirical
model that includes a chromosphere. The temperature and electron
number density for this model are plotted in Fig.~\ref{fig:FAL_C}.

Although this model does not represent the inhomogeneous and
non-equilibrium nature of the real chromosphere, it should provide a
first-order description and allow identification of the most important
mechanisms involved in the formation of the off-limb emission.
\begin{figure}
\includegraphics[width=8.8cm]{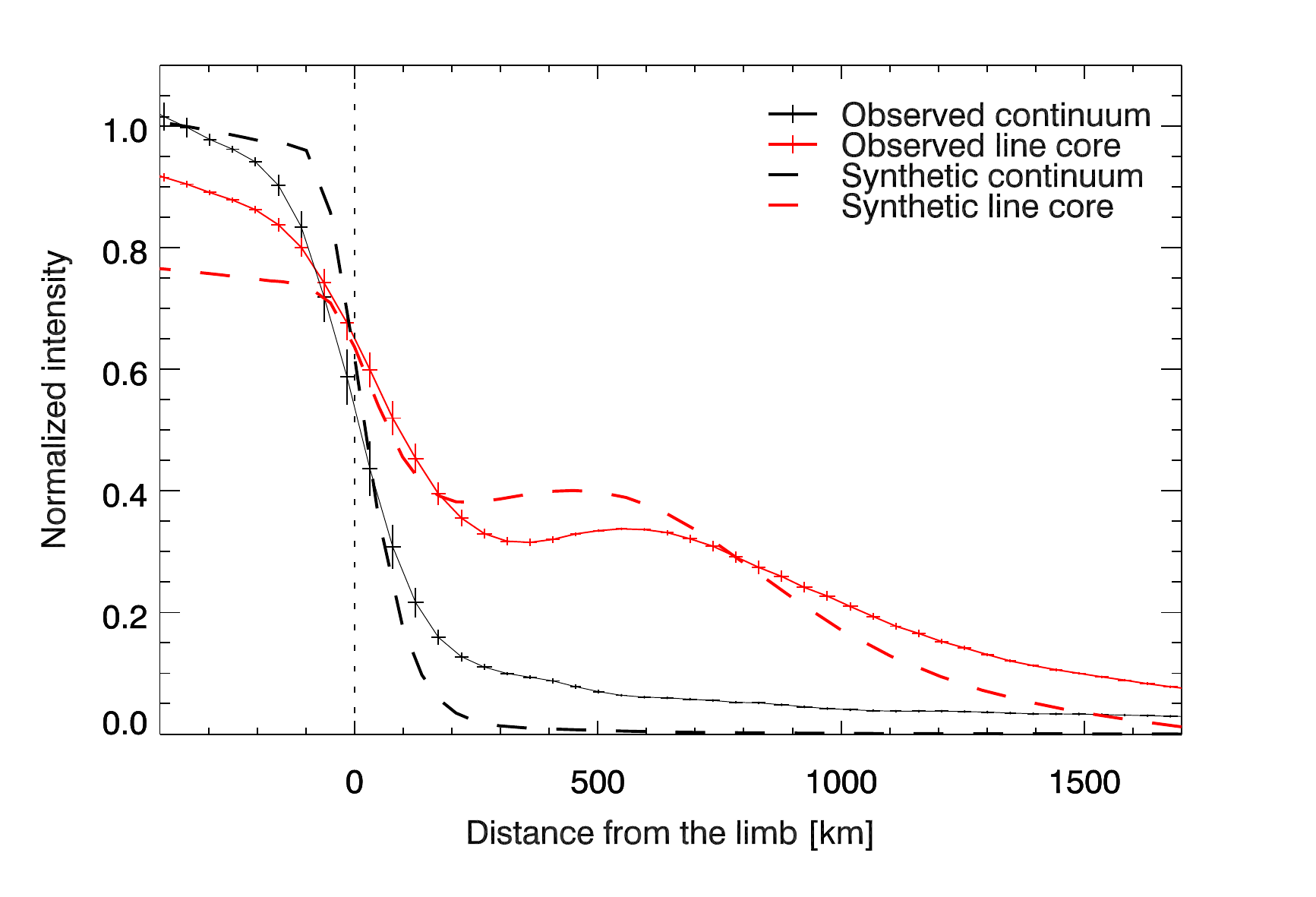}
\caption{Observed (solid) and synthetic (dashed) intensity
 profiles. The vertical and horizontal bars in the observation show the
 error from binning. The line core intensity is shown in red and the
 continuum intensity in black.}
\label{fig:obs_sim}
\end{figure}
 In order to get a realistic treatment of the coupling between oxygen and hydrogen, we also include an equation for charge conservation in our non-LTE calculations. This means that in addition to solving the statistical equilibrium equations for hydrogen and oxygen we simultaneously solve an equation for charge conservation, including non-LTE ionisation of hydrogen and oxygen and LTE ionisation for all other elements. Using the electron density as specified in FAL-C leads to inconsistencies between the \ion{H}{II} (i.e, proton) density and the electron density owing to the differences between RH and the code that was used to compute the FAL-C model.
 The final electron number density
(shown as a dashed line in Fig.~\ref{fig:FAL_C}) in
the atmospheric model differs somewhat from that in the original FAL-C model.
%______________________________________________________________
\section{Comparison with observations}
\label{sect:comp}
In order to compare the spectral modelling with our observations we have to calibrate them in space and intensity. The first calibration is to adjust the synthetic limb to the observed limb. 

For the observational data we use routines described in
\citet{2016A&A...585A.140L} 
and kindly provided by Dr. M. L{\"o}fdahl.
These are applied on the continuum image to find the
disc-centre direction and the limb position. This, together with the
known image scale, gives an $(x,y)$
or $(r,\phi)$ coordinate pair for each pixel. The portion of the
original image around the point where we want to extract the radial intensity profile
is then interpolated on an $(r,\phi)$ grid that is equidistant in $r$ with
a pixel spacing of 47~km. This is then used to produce the radial
intensity profile in Fig.~\ref{fig:obs_sim}. The inflection point of the
synthetic continuum intensity profile is used to define the zero-point
for the model output. In the following, height in the solar atmosphere and distance above the solar limb have this point as their origin.

The second calibration is to adjust the intensity. First we convolve
the synthetic spectra with a Gaussian function with the full with at half maximum (FWHM) of the
CRISP spectral instrumental profile, which is $0.096$~\AA\ at $777.1$~nm \citep{2015A&A...573A..40D}. Then we choose a point 350~km
inside the solar disc
that is present both in the observations and in the synthetic spectra
and adjust the observed intensities so 
that the two have the same continuum intensity value at this point. 
The calibration is illustrated in Fig.~\ref{fig:obs_sim}.

The model results shown in Fig.~\ref{fig:obs_sim} use the FAL-C
atmospheric model, the 54-level oxygen atom, the 6-level hydrogen
atom, and the 24-level silicon atom. In the following we
refer to this setup as the Full model, which is used as a reference for comparison with
our other numerical experiments.

The observational curves in Fig.~\ref{fig:obs_sim} come from the region at
the solar limb that is indicated on the third panel
of Fig. \ref{fig:qs}. This region clearly shows a dark gap.

The main result of the comparison is that the general shape of the observed and
simulated intensity curves are the same. We are able to reproduce the
dark gap in the line core and the off-limb emission quite well with
this 1D model. The deviations are qualitatively explained by
the finite resolution and stray light in the observations. For a quantitative comparison between observation and
model, we can calculate the ratio of intensities at the position of
maximum off-limb emission (ME) to the minimum in the dark gap (DG) for both observation and the convolved synthetic spectrum. As it is shown in Eq. \ref{equ:obs} and \ref{equ:sim}, the ratios are very close together.

\begin{equation}
\frac{I_\mathrm{obs}({\rm DG})}{I_\mathrm{obs}({\rm ME})} = 0.94
\label{equ:obs}
\end{equation}

\begin{equation} \label{equ:sim}
 \frac{I_\mathrm{sim}({\rm DG})}{I_\mathrm{sim}({\rm ME})} = 0.95
 \end{equation}
 
In the following we analyse the model in detail in order to
understand the mechanisms behind the emission and the reason for the
dark gap.

\section{Understanding the model}
\label{Sect:understand}

\begin{figure} [ t]
\includegraphics[width=8.8cm]{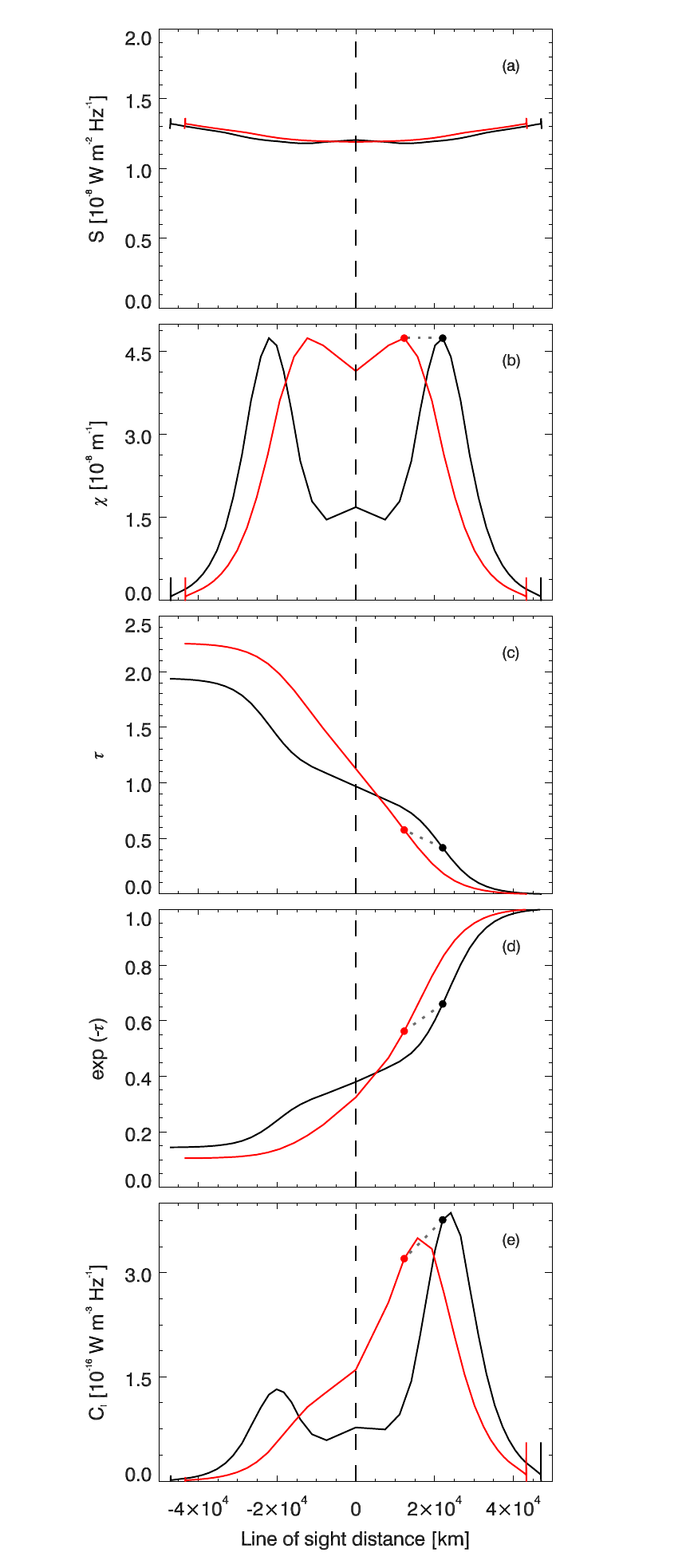}
\caption{Formation of the dark gap. The black curves show values for a ray that gives rise to the
 dark gap while the red curve represents the maximum off-limb emission. Panels, from top to bottom: source function, extinction coefficient, radial optical depth $\tau$, $\mathrm{e}^{-\tau}$, and the contribution function $C_I$, all as functions of geometrical length along the ray. The dashed lines connect points with the same geometric vertical height in the atmosphere. }
\label{fig:CF}
\end{figure}

In order to understand the nature of the off-limb emission, we have
plotted the contribution function to intensity $C_I$ and its associated variables in Fig.~\ref{fig:CF}. They are related to the emerging intensity $I$ according to Eq. \ref{eq:cf}:

\begin{equation}\label{eq:cf}
I = \int C_I  \,\mathrm{d} s = \int \chi \, S \, \mathrm{e}^{-\tau} \, \mathrm{d}s
\end{equation}

where $\chi$ is the extinction coefficient, $S$ is source function, $\tau$ the optical depth, and $s$ the geometrical distance.

\begin{figure}
\includegraphics[width=8.8cm]{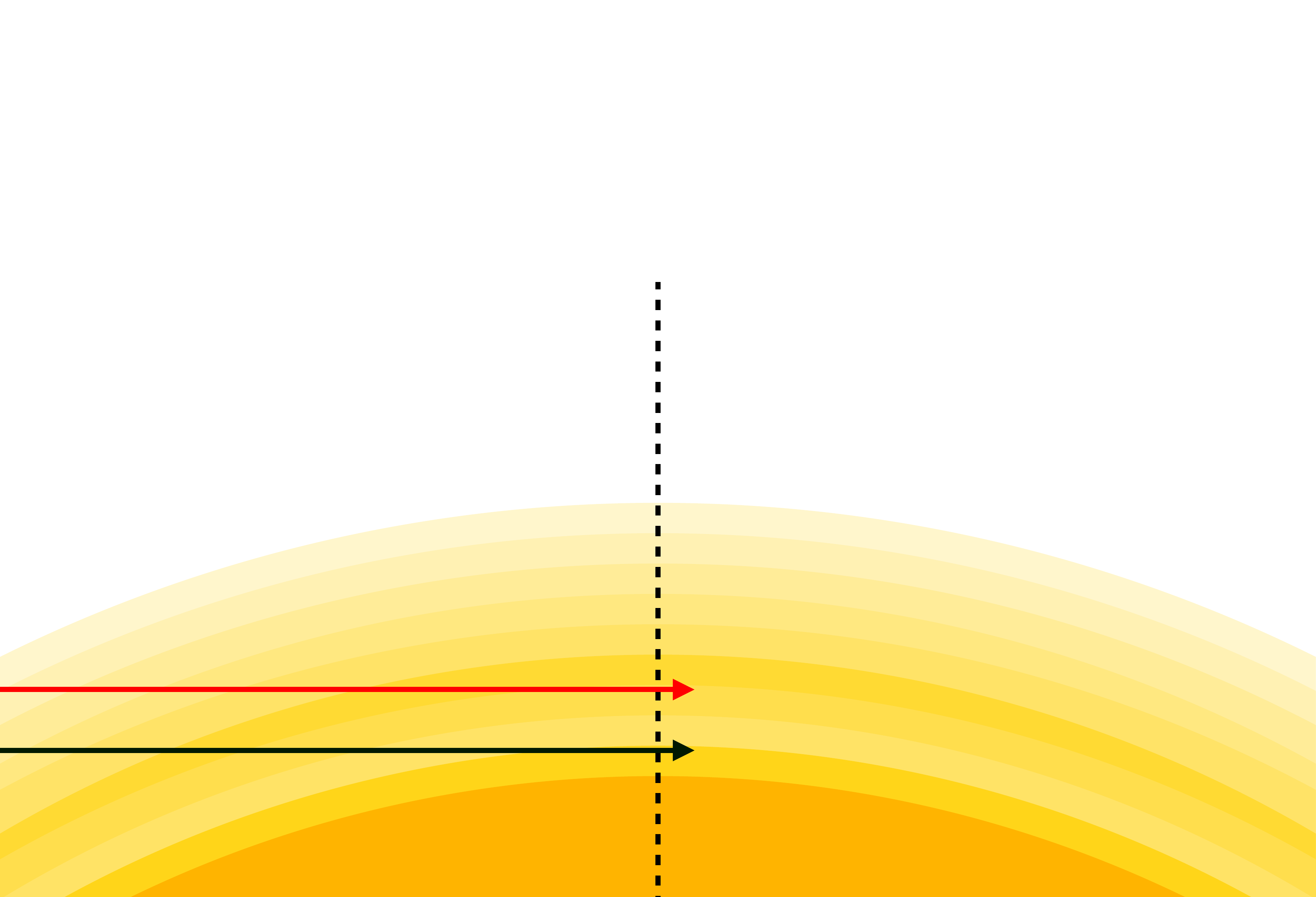}
\caption{This sketch shows the radial opacity and where the red and
 black rays go through the atmosphere. Colour-coding as in Fig.~\ref{fig:CF}.}
\label{fig:dg_sketch}
\end{figure}

Panels (a), (b), (c), (d), and (e) show the source function, the extinction coefficient, the optical depth, $\exp(-\tau)$ and the contribution function respectively. The black curves represent the ray passing through the observed dark gap towards the observer and the red curves show the ray that is passing through the point where the observed off-limb emission has its local maximum outside the dark gap. Figure \ref{fig:dg_sketch} is a sketch illustrating the geometry of the two rays.

Panel (a) of Fig.~\ref{fig:CF} shows that the source function is almost constant along both paths and that it is almost identical for the two cases. In contrast, the extinction coefficient shows a clear difference between the two rays. The black curve representing the ray passing the dark gap, shows a strong dip as it passes closest to the solar disk.
The optical depth is the extinction coefficient integrated along the
line of sight. Panel (c) shows the optical depth for the two rays. The 
dip in opacity experienced by the dark-gap ray makes the total optical
depth along that ray smaller than for the maximum-intensity ray --
even though its geometrical path length is longer and it has a larger maximum in the contribution function (panel (e)). For two homogeneous
slabs with constant and equal source functions, we expect the relation between
outgoing monochromatic intensities to be $I_1/I_2 = (1-\mathrm{e}^{-\tau_1}) /
(1-\mathrm{e}^{-\tau_2})$. In our case, panel (c) in Fig.~\ref{fig:CF}
shows that $\tau_1 = 1.93$ and $\tau_2 = 2.25$ giving $I_1/I_2 = I_{\rm  DG}/I_{\rm ME}= 0.95$
which is equal to the value from the Full model calculation.

In summary, the off-limb optical depth in the dark gap at the line core of \ion{O}{I}~7772 is of the order of unity and its line source function is more or less constant over the relevant interval. Therefore it is primarily variations in opacity that cause variations in the observed intensity. 

 In the following we discuss the processes that control the
 populations of the levels involved in the line.

\subsection{Line emission mechanism}
\begin{figure} [t]
\includegraphics[width=8.8cm, trim={0 3.74cm 0 0},clip]{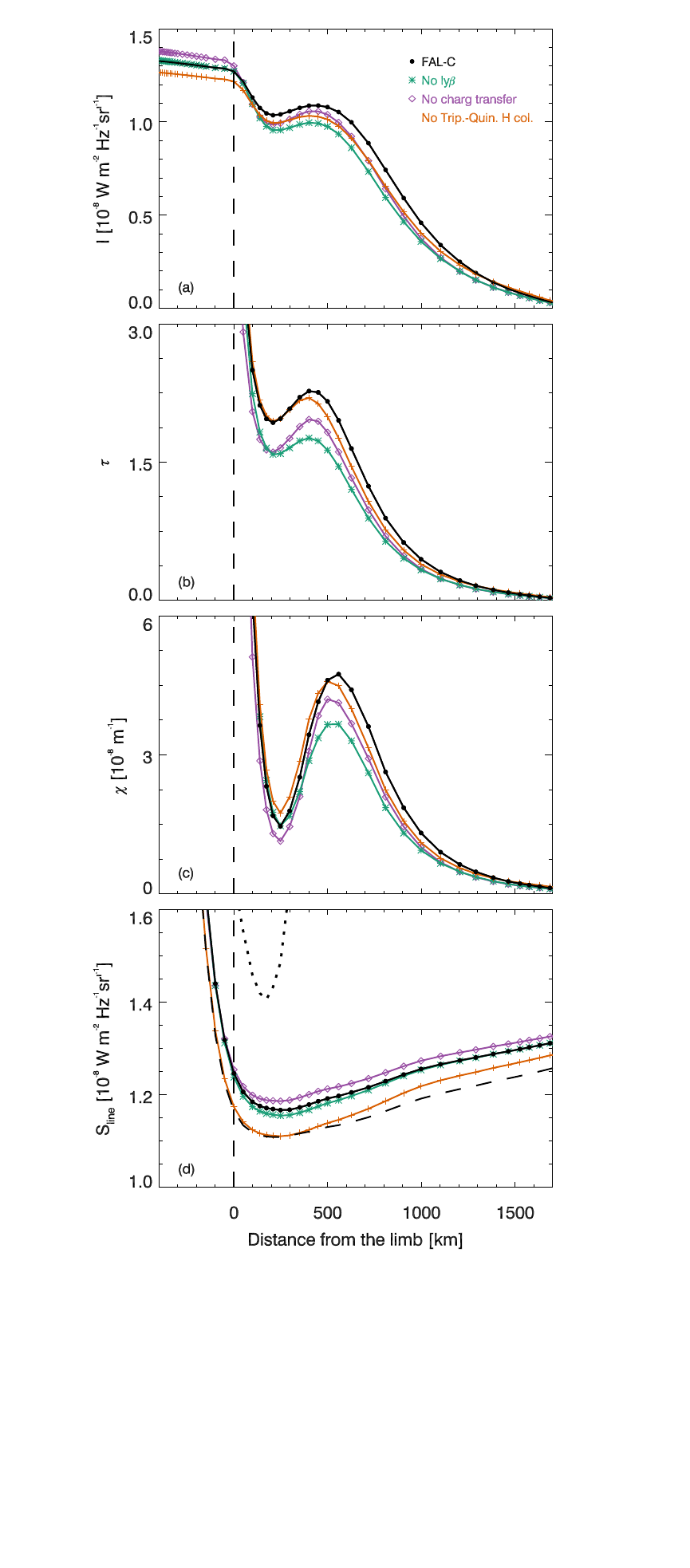}
\caption{The Full model in FAL-C (black lines) compared with results where the
 atomic processes involving hydrogen have been perturbed. (a)
 Monochromatic intensity at line core, (b) tangential $\tau$, (c)
 opacity, and (d) monochromatic line source function. The dashed line
 in panel (d) shows the mean intensity $\bar J$ for the Full model and the dotted line represents the Planck function.}
\label{fig:H}
\end{figure}
%%%%%%%%

To better understand how the off-limb emission is formed, we discuss the behaviour of the black curves in Fig.~\ref{fig:H}. These curves represent monochromatic quantities in the Full model as a function of vertical height with the zero-point corresponding to the apparent limb in the continuum as defined in Sect.~\ref{sect:comp}. The coloured curves are perturbations of the model that will be discussed later. The panels show the intensity (a), tangential optical depth (b), opacity (c) and line source function (d).

A comparison between panels (a-c) shows that the intensity, the tangential optical depth and the opacity have the same general shape, vindicating the view that it is mainly the opacity that modulates the off-limb intensity. A small shift between the opacity and optical depth curves is explained by the spherical geometry -- a ray passing through the opacity maximum when closest to the surface will pass through higher layers with lower opacity for the rest of its path.

Panel (d) of Fig.~\ref{fig:H} shows the line source function. The dashed line in this plot is the mean intensity as averaged over the line profile (${\bar{J}}$) for the Full model. 
The line source function is close to ${\bar{J}}$ and displays the same behaviour with height. 
We interpret this as a continuation of the well-known photospheric
behaviour of this line source function, where it is essentially a
two-level-atom source function $S^L = (1-\varepsilon) \bar J + \varepsilon B_\nu$.

At these heights the collisional contribution ($\varepsilon B$) is very small. 
Thus we can regard the line source function as a two-level-atom source function, being dominated by scattering, with some perturbation in the form of an additional emissivity. $\bar J$ is approximately constant with height over the region of interest. This reinforces the view that it is the opacity that needs to be understood and we use the remainder this section to discuss how level 5, the lower level of the 7772~\AA\ line, becomes populated.

We firstly note that collisional excitation from the ground state is
unlikely to be important for populating the excited levels.
LTE level populations (which is what we would get if collisions were
dominating) are far below the results from our Full model.

\begin{figure}[t]
\includegraphics[width=8.8cm]{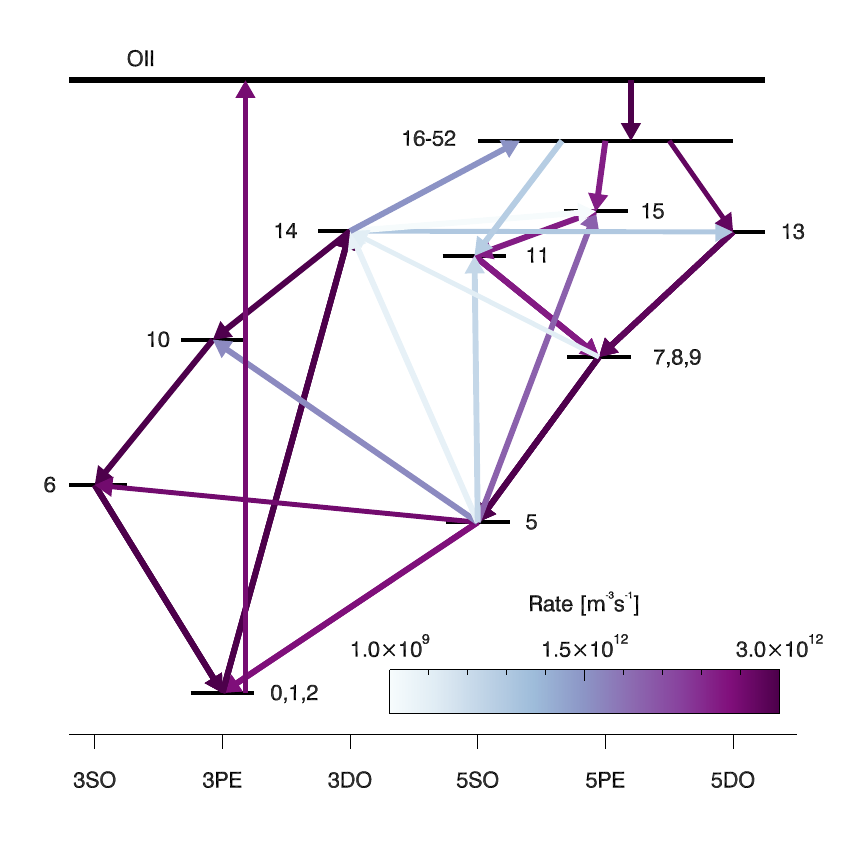}
\caption{Net rates visualised in the term diagram for the maximum
 off-limb opacity height. The colours of the arrows indicate the
 logarithm of the absolute value of the net rate with light blue the lowest rates and dark purple the 
  highest net rates.}
\label{fig:net_rate}
\end{figure}

\begin{figure}[ t]
\includegraphics[width=8.8cm]{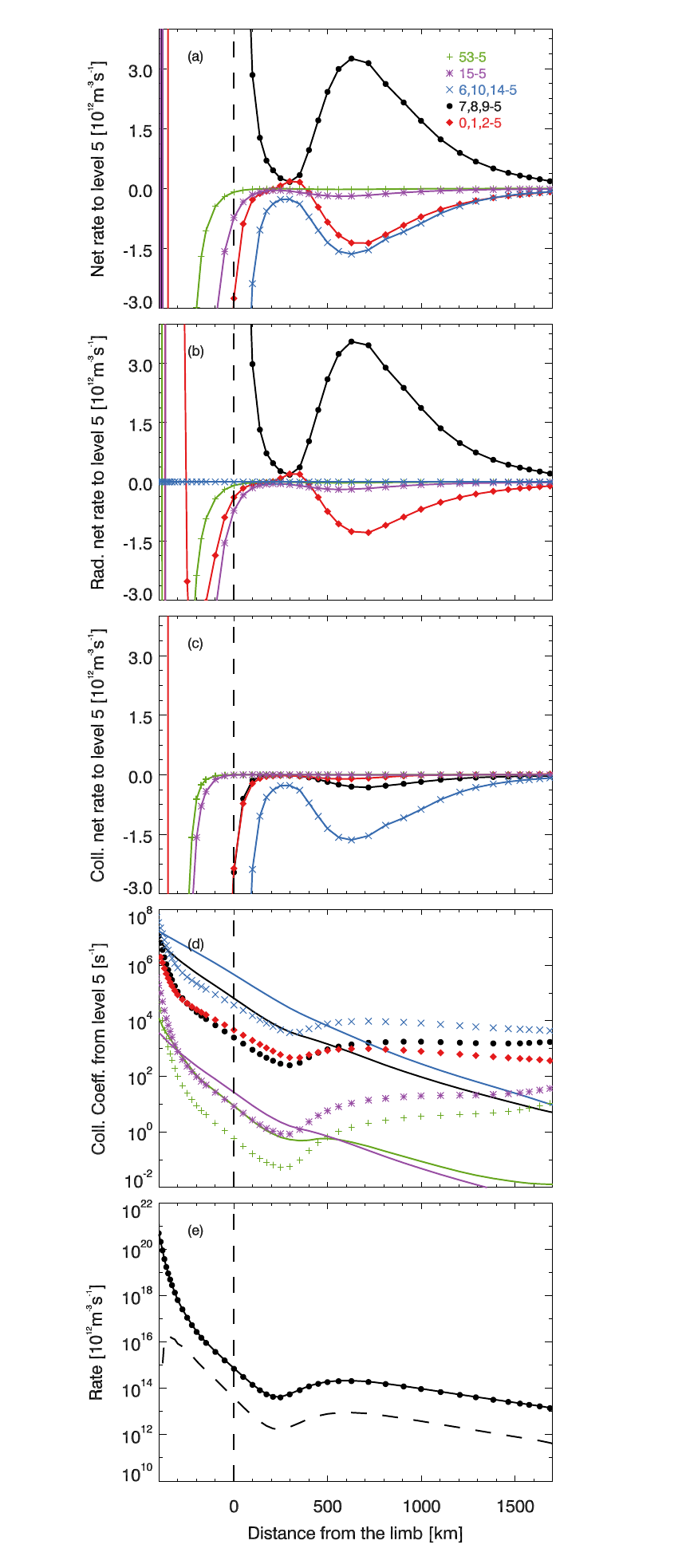}
\caption{Panels from top to bottom represent (a) net rates,
 (b) radiative net rates, and (c) collisional net rates. Panel (d) shows
 the collisional rate per atom in level 5 with hydrogen collisions
 being represented by connected symbols. Panel (e) displays the 7772~\AA\ line's downward rate in solid, upward rate with symbols and the net rate with the dashed line.}
\label{fig:CR}
\end{figure}

Panels (a-c) of Fig.~\ref{fig:CR} display net rates for transitions and groups of transitions 
that populate (positive values) or depopulate (negative values) level 5.
Remember that we are mainly concerned with the height around 500~km where the ray of the maximum emission passes.
Fig.~\ref{fig:net_rate} visualises the net rates for that height in the shape of arrows where the arrow head signifies the direction of the net rate and the colour signals its absolute value.

Inspection of the plots shows that level 5 is populated by radiative
transitions from higher levels. The 7772~\AA\ line is part of a
recombination flow that is enhanced by transitions from the triplet
system to higher levels of the quintet system (levels 13 and above).
This also results in the increased line source function relative to 
$\bar J$ as seen in panel (d) of Fig.~\ref{fig:CR}. The relative
increase is about 10\%.

Level 5 is depopulated by radiative and collisional transitions to the ground state (levels 0, 1, 2) and collisionally to excited levels in the triplet system (levels 6, 10, 14).

Panel (d) of Fig.~\ref{fig:CR} shows rate coefficients for collisions
with electrons (unconnected symbols) and neutral hydrogen atoms
(solid lines). This is done for the same (compound) transitions
as for the net rate plots. The rates are per atom in the level 5
state. At the height of interest (around 500 km), hydrogen and electron
collisions are approximately equally important. 

In our model we have the new collisional rates for hydrogen, calculated in detail for levels up to level 20.
This means that all curves in panel (d) are computed in this way except transitions to the continuum (black plus symbols) which
retain the Drawin-based estimate.
It is interesting to note that, in the region of interest around 500~km over the limb, the electron rates and hydrogen rates are
of the same order of magnitude. The lower hydrogen cross-sections are compensated for by the preponderance of hydrogen atoms
over electrons.
The exception is the 1356~{\AA} intersystem transition (0,1,2-5) where the hydrogen collisions are essentially zero.

In Sect.~\ref{sec:heff} we look further into the role that hydrogen plays for the relevant oxygen transmissions.

The final panel (e) of Fig.~\ref{fig:CR} displays the upward, downward, and net radiative rates in the
7772~\AA\ transition. We see that the net rate is small
compared to the upward and downward rates. This shows
that the line is mainly scattering radiation from the photosphere.

\subsection{Effects of coupling with hydrogen} \label{sec:heff}

The \ion{O}{I} radiative transfer problem couples with hydrogen in interesting ways that are worthy of a separate discussion.

In the current study, we can divide the effects of hydrogen into three main processes: Charge transfer between neutral oxygen and ionised hydrogen, Ly$\beta$ pumping/suction, and inelastic collisions involving neutral hydrogen atoms.

We investigate the importance of these processes on the formation of the 7772~\AA\ off-limb emission through numerical experiments. The results of these are represented by coloured curves in Fig.~\ref{fig:H} where they are compared with the results from the Full model.

The close correspondence between the ionisation potential of oxygen
(13.62 eV) and hydrogen (13.60 eV) makes collisional charge transfer
between them likely. The result is that the ionisation ratio of oxygen
is strongly coupled to that of hydrogen. \citet{1971ApJ...166...59F}
showed that in the limit of detailed balance the ionisation ratios of
oxygen and hydrogen are given by

 \begin{equation}
 \label{equ:ratio}
 \frac{n(\rm O)}{n(\rm O^+)} = \frac{9}{8}\frac{n({\rm
    H})}{n(\rm H^+)}
 \end{equation}

The modelling of \citet{1993ApJ...402..344C} found this to be the case in the solar
atmosphere. Our modelling did not achieve this ratio unless we updated
both the electron and the proton density so that everything was consistent. That is one reason why we
chose to treat hydrogen as an active species together with oxygen (and
silicon). 

The purple curves in Fig.~\ref{fig:H} show what happens when the charge-transfer transitions are removed from the model. 
The effect is that the recombination cascade dries up, leading to a decrease in the population of level 5 and thus the line opacity.
The source function increases somewhat due to the weakening of the photospheric absorption line (leading to increasing $\bar J$), but this
is not enough to dominate over the opacity effect.

%\comment{2. Ly-beta. Quenching it leads to lower opacity. Ly-beta pumping spills over to the quintet system via collisional coupling from level 14.Source function drops somewhat. That is a photospheric effect.}

To examine the effect of Ly$\beta$ on the formation of this line, we remove the Ly$\beta$ radiative transition from our hydrogen atom. In order to keep the hydrogen populations at the correct values, we insert the values from the Full model in our atmosphere model and run hydrogen as a background element. The green curve in Fig. \ref{fig:H} shows the results when Ly$\beta$ pumping is turned off. 

The result is that the off-limb intensity decreases by about 10\% as the line opacity also decreases. 

%\comment{3.2 Removing them in triplet-quintet transitions lower opacity somewhat because the effects fo the Ly-beta pumping are quenched.
%WHY does the source function drop? It drops to J. But that is J in the Full model. Clearly this is a photospheric effect. 
%How can the photospheric effect be so strong?}

The coupling between the triplet and quintet systems is of interest
because it determines how effectively the Ly$\beta$ pumping is
transmitted to the quintet levels. Removing the hydrogen collisions
between the triplet and quintet levels results in the orange curves in
Fig.~\ref{fig:H}. This perturbation gives only a small decrease in the
level 5 population and thus the line opacity. The effect is stronger
on the photospheric line which gets stronger, causing a depression of
the line source function in higher layers by lowering $\bar J$ there.

We note that the new rates for hydrogen collisions significantly
decrease the coupling between the excited levels of the triplet and
quintet system. When using collisional rates calculated with the
Drawins recipe, the importance of the Ly$\beta$ pumping on the triplet
system is much greater. For the current study of off-limb emission, this is the most
important impact of the new rates.

\subsection{Dependence on atmospheric structure}

\begin{figure}[t]
\includegraphics[width=8.8cm, trim={0 3.74cm 0 0},clip]{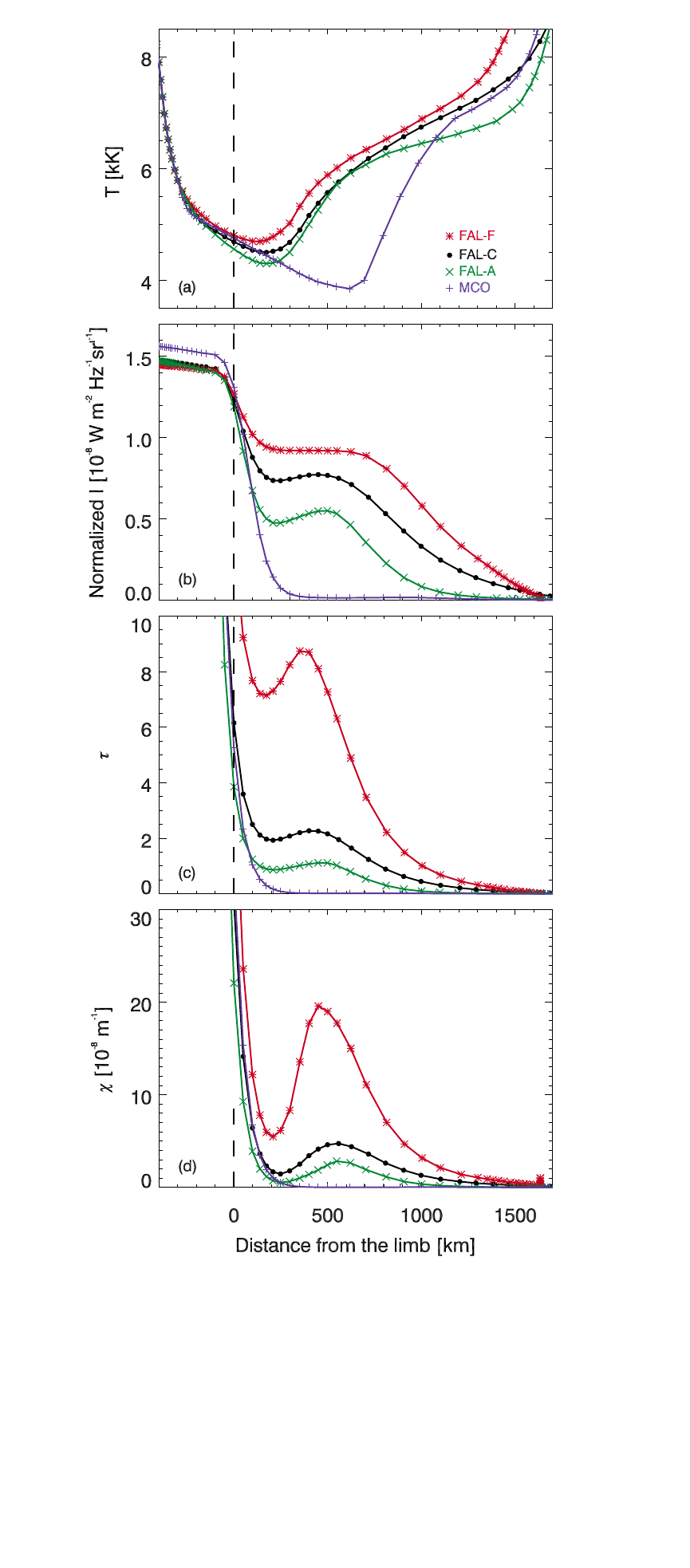}
\caption{Similar plots as in Fig.~\ref{fig:H} (but with temperature
 instead of line source function) for different atmospheric models.}
\label{fig:atmos}
\end{figure}

To explore the effects of atmospheric structure on the off-limb
emission, we repeat the same calculation with different atmospheric
models from the FAL family \citep{1993ApJ...406..319F}. The results are plotted in Fig. \ref{fig:atmos}.

Panel (d) of this figure shows the temperature structure of the
models. The plots have been treated in exactly the same way as for the
FAL-C model so that the limb position has been individually adjusted.

FAL-F represents bright network elements and has higher
temperatures than FAL-C. Here we can see that the dark gap completely disappears
as the line becomes optically thick. The off-limb emission is also boosted by the
fact that the effective formation height moves upwards where the source
function is greater.

FAL-A is a little cooler than FAL-C. The decreased opacity leads to lower emission.

MCO \citep{1997ApJ...484..960G} is made to fit
observations of infrared CO lines and has much lower temperatures than
FAL-C. The opacity and the tangential $\tau$ in this model stay close to
zero and so does the off-limb emission.

This range of models shows how the observed variations in intensity and
clarity of the dark gap is well within the range of reasonable
variations of atmospheric structure.

\citet{2015ApJ...813...34L} explain how the intensity of the 1356~\AA\ line emission is
proportional to the square of the electron density -- an effect of the
recombination cascade being dependent on the proton density through
the charge-transfer coupling between oxygen and hydrogen and the
importance of recombination. This emission is
formed under optically thin conditions at a somewhat greater height
than when the 7772~\AA\ emission peaks. The upper level of the
1356~\AA\ line is our level 5 which sets the opacity in the
7772~\AA\ line. While the maximum of the 7772~\AA\ off-limb emission is
predicted to be formed lower than that of 1356~\AA, the difference is
about 350~km in our model, we expect that the situation is
similar. Thus we can expect that the 7772~\AA\ line opacity also will vary as the
square of the electron density, at least approximately. The dark gap does indeed correspond to
the minimum of $n_e$ corresponding to the temperature minimum.

\section{Conclusions}

We have found that the spatial structures seen in the off-limb emission of the
7772~{\AA} line -- notably the dark gap -- are formed by opacity
variations. 

We find that the lower level population and thus the opacity is mainly regulated through recombination cascades from \ion{O}{II}. The \ion{O}{II} population in turn is slaved to the \ion{H}{II} population through charge transfer.

The 7772~{\AA} line is mainly scattering the photospheric radiation
and so the line source function is rather constant over the region of
interest. We propose that this view is valid also for structures that
are not present in our 1D modelling but are visible in the
observational data: Spicules, prominences, and surges. 

The dark gap frequently seen over the solar limb 
is formed by the decrease in line opacity where the temperature and
the hydrogen ionisation degree (and thus the oxygen ionisation degree)
reach a minimum.
 One may say that when we are looking at the dark gap, we are actually seeing the temperature
minimum.

The success of our modelling using the 1D FAL-C atmospheric model
shows that this approach can be used to identify the relevant physical processes.

While the emission features of 1356~\AA\ and 7772~\AA\ are expected to
form at somewhat different heights, simultaneous observations of the two
lines would be an interesting consistency check on this and other
modelling efforts.

%______________________________________________________________
%______________________________________________________________
\begin{acknowledgements}
We thank Paul Barklem for providing the new hydrogen collisional cross-sections fresh from the computer. 
Pit S{\"u}tterlin is thanked for assistance during the observations.
Thanks are also due to Jaime de la Cruz Rodr{\'i}guez for answering
many questions about data reduction and to Mats L{\"o}fdahl for
lending his limb-fitting software.

The Swedish 1-m Solar Telescope is operated on the island of La Palma by the Institute for Solar
Physics of Stockholm University in the Spanish Observatorio del Roque
de los Muchachos of the Instituto de Astrof{\'i}sica de Canarias.
\end{acknowledgements}
%-------------------------------------------------------------------
\bibliographystyle{aa}
\bibliography{references}

\end{document}